\documentclass{article}

\usepackage{amssymb, amsmath, amsthm, tikz, verbatim, graphicx,lmodern,indentfirst,enumerate,color}

\usepackage[titletoc,page]{appendix}
\usepackage[margin=1.4in]{geometry}

\usepackage{pgf}

\newtheorem{thm}{Theorem}[section]
\newtheorem{lem}[thm]{Lemma}
\newtheorem{prop}[thm]{Proposition}

\theoremstyle{definition}

\theoremstyle{definition}
\newtheorem{defn}[thm]{Definition}

\theoremstyle{remark}

\numberwithin{equation}{section}

\makeatletter
\newcommand{\rmnum}[1]{\romannumeral #1}
\newcommand{\Rmnum}[1]{\expandafter\@slowromancap\romannumeral #1@}
\newcommand\restr[2]{{
  \left.\kern-\nulldelimiterspace 
  #1 
  \right|_{#2} 
  }}

\makeatother
\begin{document}

\title{A necessary and sufficient condition for the existence of chaotic dynamics in a neoclassical growth model with a pollution effect}
\author{Tomohiro Uchiyama\\
Faculty of International Liberal Arts, Soka University,\\ 
1-236 Tangi-machi, Hachioji-shi, Tokyo 192-8577, Japan\\
\texttt{Email:t.uchiyama2170@gmail.com}}
\date{}
\maketitle 

\begin{abstract}
In this paper, we study a neoclassical growth model with a (productivity inhibiting) pollution effect. In particular, we obtain a necessary and sufficient condition for the existence of a topological chaos. We investigate how the condition changes as the strength of the pollution effect changes. This is a new application of a recent result characterising the existence of a topological chaos for a unimodal interval map by Deng, Khan, Mitra (2022). 
\end{abstract}

\section{Introduction}
Our starting point of this paper is a (somewhat simplified) neoclassical growth model~\cite{Solow-Growth-QJE} in discrete time with a production lag, expressed by the following difference equation: 
\begin{equation}\label{production1}
k_{t+1}=s\cdot f(k_t)
\end{equation}
where $k_t$ is a capital-labour ratio $K_t/L_t$, $s$ is a constant saving ratio, $f$ is a per-capita production function. (We assume no population growth.) Let $f$ be the Cobb-Douglas production function:
\begin{equation}\label{production2}
f(k)=A k^\gamma \textup{ where } A >0 \textup{ and } 0 < \gamma \leq 1.
\end{equation}
From Equations (\ref{production1}) and (\ref{production2}), we obtain
\begin{equation*}\label{basicDynamics}
k_{t+1}=s A k_t^\gamma.
\end{equation*}
It is easy to see that: 1.~if $0<\gamma<1$, then $k$ converges to the steady state $k^*=(sA)^{1/(1-\gamma)}$, 2.~if $\gamma=1$, then $k$ converges to $0$, stays at $k_0$, or diverges to $\infty$ depending on the value of $sA$. The point is that, in any case, a chaotic behaviour does not emerge. (We define what we mean by a chaos below (Definition~\ref{Chaos}). There are several definitions of a chaos in literature.)  
In the rest of the paper, we assume $\gamma=1$ to simplify the argument. 

Now, following~\cite[Sec.~\Rmnum{2}.B]{Day-Irregular-AER}, we introduce a productivity inhibiting effect ("pollution effect") into our model. Then the production function becomes
\begin{equation}\label{productionPollution}
f(k)=Ak(1-k)^\beta \textup{ where } \beta\geq 0.
\end{equation}
Note that in (\ref{productionPollution}), if $k$ gets close to the maximum value (that we set $1$ to simplify the argument), the term $(1-k)^{\beta}$ gets (very) close to $0$ (if $\beta$ is large). This is what we mean by the "pollution effect". With this pollution effect, we finally obtain the difference equation we study in this paper:
\begin{equation}\label{dynamics}
k_{t+1}=f(k_t)=s A k_t (1-k_t)^\beta=\alpha k_t (1-k_t)^\beta \textup{ where } \alpha>0 \textup{ and }0\leq k_t\leq 1.
\end{equation}

Note that in (\ref{dynamics}), if $\beta=1$, then $f$ reduces to a well-studied classical unimodal map. It is known that a Li-Yorke chaos (that always comes with a cycle of period three) emerges if $3.83\leq \alpha \leq 4$, see~\cite[p.89]{Yorke-Yorke-book}.

In~\cite[Sec.~\Rmnum{2}.B]{Day-Irregular-AER}, the same equation (even with any $\gamma$ and any population growth rate) was studied, and Day showed that for any $\beta$, there exists some $\alpha$ that guarantees the existence of a Li-Yorke chaos. 
However, the proof for that result was an existential one; that does not give any clear (algebraic/numerical) relation between $\beta$ and $\alpha$. In this paper, we use algebraic argument mixed with some numerical computations with Python to investigate how a \emph{necessary and sufficient} condition (in terms of $\alpha$) for the existence of a \emph{topological chaos} (that is weaker than a Li-Yorke chaos) changes as $\beta$ ("pollution effect") changes. (See Theorem~\ref{main} for the precise statement.)

To state our main result (Theorem~\ref{main}), we need some preparation. First, we clarify what we mean by a topological chaos (or a turbulence). The following definition is taken from~\cite[Chap.~\Rmnum{2}]{Block-book}:
\begin{defn}\label{Chaos}
Let $g$ be a continuous map of a closed interval $I$ into itself. We call $g$ \emph{turbulent} if there exist three points, $x_1$, $x_2$, and $x_3$ in $I$ such that $g(x_2)=g(x_1)=x_1$ and $g(x_3)=x_2$ with either $x_1<x_3<x_2$ or $x_2<x_3<x_1$. Moreover, we call $g$ \emph{(topologically) chaotic} if some iterate of $g$ is turbulent.  
\end{defn}
It is known that a map $g$ is chaotic (in our sense) if and only if $g$ has a periodic point whose period is not a power of $2$, see~\cite[Chap.~\Rmnum{2}]{Block-book}. This implies that a map $g$ is chaotic if and only if the topological entropy of $g$ is positive, see~\cite[Chap.~\Rmnum{8}]{Block-book}. See~\cite{Ruette-book} for more characterisations of chaos. 

Second, we recall the following key result characterising the existence of a topological chaos and a turbulence~\cite[Thms.~2 and~3]{Deng-TopChaos-JET} since our main result is a direct application of this result. Let $\mathfrak{G}$ be the set of continuous maps from a closed interval $[a, b]$ to itself so that an arbitrary element $g\in \mathfrak{G}$ satisfies the following two properties:
\begin{enumerate}
\item{there exists $m\in (a,b)$ with the map $g$ strictly increasing on $[a,m]$ and strictly decreasing on $[m,b]$.}
\item{$g(a)\geq a$, $g(b)<b$, and $g(x)>x$ for all $x\in(a,m]$.}
\end{enumerate}
For $g\in \mathfrak{G}$, let $\Pi:=\{x\in [m,b]\mid g(x)\in [m,b] \textup{ and } g^2(x)=x\}$. Now we are ready to state~\cite[Thms.~2 and~3]{Deng-TopChaos-JET}:
\begin{prop}\label{ChaosThm}
Let $g\in \mathfrak{G}$. The map $g$ has an odd-period cycle if and only if $g^2(m) < m$ and $g^3(m) < \min\{x\in \Pi\}$ and the second iterate $g^2$ is turbulent if and only if $g^2(m) < m$ and $g^3(m) \leq \max\{x\in \Pi\}$. 
\end{prop}

Applying Proposition~\ref{ChaosThm} to (\ref{dynamics}) (this is a non-trivial step), we obtain our main result:
\begin{thm}\label{main}
In (\ref{dynamics}), let $\beta=1$ ($2$, $3$, or $10$ respectively). If $2<\alpha\leq 4$ ($2.250<\alpha\leq 6.750$, $2.370<\alpha\leq 9.481$, or $2.594<\alpha\leq 28.531$ respectively) then the map $f$ has the following properties:
\begin{enumerate}
\item{$f\in\mathfrak{G}$.}
\item{$f$ has an odd period cycle if and only if $3.679<\alpha\leq 4$ ($5.574<\alpha\leq 6.750$, $7.027<\alpha\leq 9.481$, or $11.795<\alpha\leq 28.531$ respectively).}
\item{The second iterate $f^2$ is turbulent if and only if 
$3.679\leq\alpha\leq 4$ ($5.574\leq\alpha\leq 6.750$, $7.027\leq\alpha\leq 9.481$, or $11.795\leq\alpha\leq 28.531$ respectively).}
\end{enumerate}
\end{thm}

In~\cite[Sec.~\Rmnum{2}.B]{Day-Irregular-AER}, Day cites~\cite{Yorke-Yorke-book} (possibly Fig.4.3 on p.88)
and states that $3.57\leq \alpha \leq 4$ is a sufficient condition for the existence of a Li-Yorke chaos. Our result shows that this is wrong since a topological chaos is weaker than a Li-Yorke chaos.


In the next section, we show that the upper bound for $\alpha$ for each $\beta$ in Theorem~\ref{main} comes from the simple requirement that $0\leq k\leq 1$. So the upper bound is nothing to do with the existence of a chaos. The real meat is in the lower bound. The general pattern we can see from Theorem~\ref{main} is that if $\beta$ ("pollution effect") becomes large, then the lower bound for $\alpha$ gets loose. (The range of $\alpha$ gets wider.) In other words, to generate a chaos, we need either a large $\alpha$ (this controls a vertical stretching of the graph of $f$) or a large $\beta$ (strong pollution effect), but not necessarily both. (Of course, we need some pollution effect: no chaos emerges with zero pollution effect as we have shown above.)

\section{Proof of Theorem~\ref{main}}
\subsection{General argument}
We keep the same notation from Introduction such as $\mathfrak{G}$, $m$, $\Pi$ etc. First, we have $f'(k)=\alpha(1-k)^{\beta-1}(1-k-\beta k)$. So setting $f'(k)=0$, we obtain $k=1/(1+\beta)$ or $k=1$ if $\beta>1$. Since $f'(k)>0$ for $0<k<1/(1+\beta)$ and $f'(k)<0$ for $1/(1+\beta)<k<1$, $f$ is unimodal on $[0,1]$ and takes its maximum at $k=1/(1+\beta)$. Set $m:=1/(1+\beta)$ and $I:=[0,1]$.  
Since we want $f$ to be a map from $I$ to itself, we need $0\leq f(m) \leq 1$. Solving this, we obtain 
\begin{equation}\label{condition1}
\alpha \leq (\beta+1)\left(\frac{\beta+1}{\beta}\right)^{\beta}.
\end{equation}

Next, since we want to push $f$ into $\mathfrak{G}$, we need $f(k)>k$ for any $k\in (0,m]$. So in particular, we need $f(m)>m$. Solving this, we obtain 
\begin{equation}\label{condition2}
\left(\frac{\beta+1}{\beta}\right)^\beta < \alpha.
\end{equation}
Combining (\ref{condition1}) and (\ref{condition2}) yields
\begin{equation}\label{condition3}
\left(\frac{\beta+1}{\beta}\right)^\beta < \alpha \leq (\beta+1)\left(\frac{\beta+1}{\beta}\right)^{\beta}.
\end{equation}
Note that $f(m)>m$ implies that $f(k)>k$ for any $k\in (0,m]$. This is because $f(k)-k=k(\alpha(1-k)^\beta-1)>0$ is equivalent to $\alpha(1-k)^\beta-1>0$, but $\alpha(1-k)^\beta-1\geq\alpha(1-m)^\beta-1$ for any $k\in (0,m]$. An easy calculation shows that $\alpha(1-m)^\beta-1>0$ is equivalent to (\ref{condition2}). Other conditions for $f$ to be in $\mathfrak{G}$ are clearly satisfied, so we have shown that:
\begin{lem}\label{firstLem}
If $\left(\frac{\beta+1}{\beta}\right)^\beta < \alpha \leq (\beta+1)\left(\frac{\beta+1}{\beta}\right)^{\beta}$, then $f\in \mathfrak{G}$.
\end{lem}

\begin{figure}[h]
	\begin{center}
    	\scalebox{.4}{\input{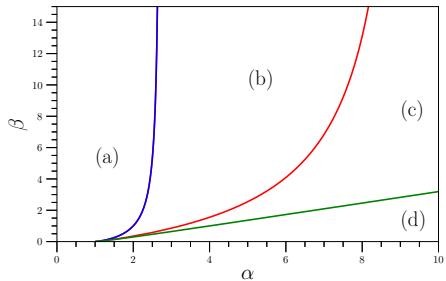}}
	\end{center}
    \caption{$(\alpha,\beta)$ with (\ref{condition3}) and (\ref{squaredCondition})}\label{fig1}
\end{figure}

Next, we consider the condition $f^2(m)<m$ in Proposition~\ref{ChaosThm}. By an easy computation, we see that this is equivalent to
\begin{equation}\label{squaredCondition}
1-\alpha^2 \left(\frac{\beta}{\beta+1}\right)^\beta\left(\frac{-\alpha\left(\frac{\beta}{\beta+1}\right)^\beta+\beta+1}{\beta+1}\right)^\beta>0
\end{equation}
Plotting $(\alpha,\beta)$ satisfying (\ref{squaredCondition}), we obtain Figure~\ref{fig1}.
In Figure~\ref{fig1}, region (b)+(c) represents all $(\alpha,\beta)$ satisfying (\ref{condition3}), region (a)+(c)+(d) represents all $(\alpha,\beta)$ satisfying (\ref{squaredCondition}). Thus, in view of Proposition~\ref{ChaosThm}, we need to choose $(\alpha, \beta)$ in region (c) to generate a chaos. The other condition in Proposition~\ref{ChaosThm}, that is $f^3(m) < \min\{x\in \Pi\}$, turned out to be too hard to express algebraically in terms of arbitrary $\alpha$ and $\beta$. The last thing we can do here is to compute fixed points of $f$; solving $f(k)=k$, we obtain $k=0, 1-(1/\alpha)^{1/\beta}$. Set $z:=1-(1/\alpha)^{1/\beta}$. It is clear that $z\in \Pi$. In the following, we choose some particular $\beta$, and express the necessary and sufficient condition to generate a chaos in terms of $\alpha$.

\subsection{$\beta=1$ case}
This case was considered in~\cite[Sec.~3.1]{Deng-TopChaos-JET}, but we redo it here since we use a different and more direct argument. For other cases, arguments are similar. 

First, (\ref{condition3}) gives $2<\alpha\leq 4$. By Lemma~\ref{firstLem}, if $2<\alpha\leq 4$ then $f\in \mathfrak{G}$. Now, by (\ref{squaredCondition}), the condition $f^2(m)<m$ translates to $f^2(1/2)=0.25\alpha^2(1-0.25\alpha)<1/2$. Solving this (numerically), we obtain $3.236<\alpha$. Next, we need to find a explicit formula for the set $\Pi$. Solving $f^2(k)=k$, we obtain $k=0$, $(\alpha-1)/\alpha$ (two fixed points of $f$), $(\alpha-\sqrt{\alpha^2-2\alpha-3}+1)/(2\alpha)$, and $(\alpha+\sqrt{\alpha^2-2\alpha-3}+1)/(2\alpha)$ (two period two points of $f$). Since $\alpha>3.236$ (in particular $\alpha>3$), both of the period two points of $f$ are real. Now we show that $\Pi=\{z\}$. To do this, it is enough to show that $(\alpha-\sqrt{\alpha^2-2\alpha-3}+1)/(2\alpha)<1/2$. Solving this, we obtain $\alpha>1+\sqrt{5} \approx 3.236$ (we know that this holds). So, $\Pi=\{z\}$. Now the condition $f^3(m)<\min\{k\in \Pi\}$ simplifies to $f^3(m)<z$. Solving this numerically, we obtain $\alpha>3.679$. Since we have $2<\alpha\leq 4$, combining these, we obtain $3.679<\alpha\leq 4$.  
 
\subsection{$\beta=2$ case}
\begin{figure}[h]
	\begin{center}
    	\scalebox{.4}{\input{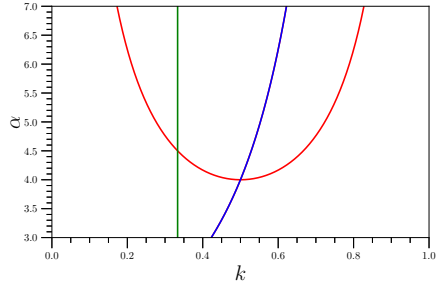}}
	\end{center}
    \caption{$(k,\alpha)$ with $\beta=2$ and $f^2(k)=k$}\label{fig2}
\end{figure}

By the same argument as the $\beta=1$ case, we need $2.250<\alpha\leq 6.750$ to have $f\in \mathfrak{G}$. 
The condition $f^2(m)<m$ is equivalent to $-4\alpha^2(4\alpha-27)^2/6561 +1 >0$. Solving this (and combining with $2.250<\alpha\leq 6.750$), we obtain $4.500 < \alpha \leq 6.750$. It is clear that the fixed points of $f$ is $k=0$ and $1-\sqrt{1/\alpha}$. Solving $f^2(k)=k$ (that is equivalent to $k\left(\alpha^2(k-1)^2(\alpha k(k-1)^2-1)^2-1\right)=0$), and plotting $(k,\alpha)$ pairs satisfying the last equation gives Figure~\ref{fig2}. In Figure~\ref{fig2}, the blue curve gives the set of nontrivial fixed points of $f$, the red curve gives the set of period two points of $f$, and the green line is the $k=m$ line. A numerical computation shows that if $\alpha>4.500$, then the left branch of the red line sits to the left of the green line. This shows that $\Pi=\{z\}$.  
Now we consider the condition $f^3(m)<z$, and (combining all the other conditions) obtain $5.574<\alpha\leq6.750$. 

\subsection{$\beta=3$ case}
\begin{figure}[h]
	\begin{center}
    	\scalebox{.4}{\input{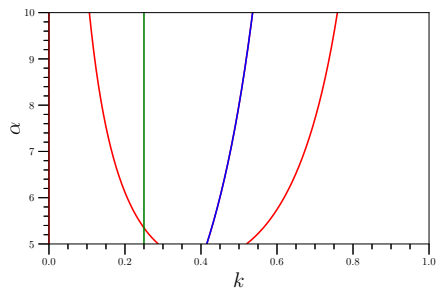}}
	\end{center}
    \caption{$(k,\alpha)$ with $\beta=3$ and $f^2(k)=k$}\label{fig3}
\end{figure}

First, to push $f$ into $\mathfrak{G}$, by the same argument as above, we need $2.370<\alpha\leq9.481$. Second, we see that $f^2(m)<m$ is equivalent to $27\alpha^2(27\alpha-256)^3/1073741824 +1 > 0$. From this, we obtain $5.347<\alpha\leq 9.481$. Third, $f^2(k)=k$ is equivalent $k(\alpha^2(k-1)^3(\alpha k (k-1)^3+1)^3+1)=0$. Now, solving the last equation gives Figure~\ref{fig3}. By the same argument as that for the last case, by a numerical computation, we have that if $\alpha>5.347$, then $\Pi=\{z\}$. 
Now, solving the condition $f^3(m)<z$ (and combining all the other conditions), we obtain $7.027<\alpha\leq 9.481$. 

\subsection{$\beta=10$ case}
\begin{figure}[h]
	\begin{center}
    	\scalebox{.4}{\input{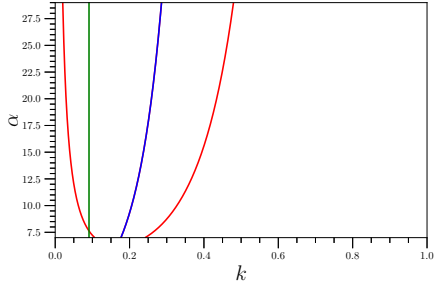}}
	\end{center}
    \caption{$(k,\alpha)$ with $\beta=10$ and $f^2(k)=k$}\label{fig4}
\end{figure}

First, we need $2.594<\alpha\leq 28.531$ to push $f$ into $\mathfrak{G}$. Next, solving $f^2(m)<m$, we obtain $7.626<\alpha\leq 28.531$. Now, from $f^2(k)=k$, that is equivalent to $a^2 k(1-k)^{10}(-a k(1-k)^{10}+1)^{10}-k=0$, we obtain Figure~\ref{fig4}. 
By a numerical computation, we obtain that if $\alpha>7.626$ then $\Pi=\{z\}$. Finally, solving $f^3(m)<z$ (and combining all the other conditions), we obtain $11.795<\alpha\leq 28.531$. 

\subsection{Conclusion}
It is clear that $\max\{k\in \Pi\}=\{z\}$ for all the cases we consider. Therefore, in view of Proposition~\ref{ChaosThm}, we have proven Theorem~\ref{main}.

\bibliography{econbib}

\begin{thebibliography}{1}

\bibitem{Block-book}
L.S. Block and W.A. Coppel.
\newblock {\em {Dynamics in One Dimension}}.
\newblock Springer, Berlin, 1992.

\bibitem{Day-Irregular-AER}
R.H. Day.
\newblock Irregular growth cycles.
\newblock {\em Am. Econ. Rev}, 72:406--414, 1982.

\bibitem{Deng-TopChaos-JET}
L.~Deng, M.A. Khan, and T.~Mitra.
\newblock Continuous unimodal maps in economic dynamics: On easily verifiable
  conditions for topological chaos.
\newblock {\em J. Econ. Theory}, 201, 2022.
\newblock Article 105446.

\bibitem{Ruette-book}
S.~Ruette.
\newblock {\em {Chaos on the Interval}}.
\newblock American Mathematical Society, Providence, 2017.

\bibitem{Solow-Growth-QJE}
R.M. Solow.
\newblock Contribution to the theory of economic growth.
\newblock {\em Q. J. Econ}, 70:65--94, 1956.

\bibitem{Yorke-Yorke-book}
J.A. Yorke and E.~Yorke.
\newblock Chaotic behavior and fluid dynamics.
\newblock 1985.
\newblock in H.L. Swinney and J.P. Gollub, eds., Hydrodynamic Instabilities and
  the Transition to Turbulence.

\end{thebibliography}

\end{document}